\documentclass[letterpaper, 10 pt, conference]{ieeeconf}  
\IEEEoverridecommandlockouts                              

\usepackage{cite}
\usepackage{url}
\usepackage{amsmath}
\usepackage{amssymb}
\usepackage{dsfont}
\usepackage{bm}
\usepackage[caption=false,font=footnotesize]{subfig}
\usepackage{siunitx}
\usepackage{algorithm}
\usepackage{graphicx}
\usepackage{epstopdf}
\usepackage{booktabs}
\usepackage{diagbox}
\usepackage{enumerate}
\usepackage{color}
\graphicspath{{figures/}}
\usepackage{multirow}

\newcommand{\mathletter}[1]{%
	\expandafter\newcommand\csname b#1\endcsname{\mathbb #1}
	\expandafter\newcommand\csname c#1\endcsname{\mathcal #1}
	\expandafter\newcommand\csname f#1\endcsname{\mathfrak #1}
	\expandafter\newcommand\csname til#1\endcsname{\widetilde #1}
	\expandafter\newcommand\csname ha#1\endcsname{\widehat #1}
	\expandafter\newcommand\csname bf#1\endcsname{\bf #1}
}%
\def\mathletters#1{\mathlettersB #1,,}
\def\mathlettersB#1,{\ifx,#1,\else\mathletter #1\expandafter\mathlettersB\fi}
\mathletters{A,B,C,D,E,F,G,H,I,J,K,L,M,N,O,P,Q,R,S,T,U,V,W,X,Y,Z}

\def\bee{\begin{equation}}
	\def\ene{\end{equation}}
\def\beq{\begin{eqnarray}}
	\def\enq{\end{eqnarray}}
\def\bmatri{\begin{bmatrix}}
	\def\ematri{\end{bmatrix}}

\newtheorem{lemma}{Lemma}

\newtheorem{remark}{Remark}
\newtheorem{prop}{Proposition}

\title{\bf Coordinate-free Isoline Tracking in Unknown  2-D Scalar Fields}

\author{Fei~Dong, Keyou~You,~\IEEEmembership{Senior Member,~IEEE}
	\thanks{*This work was supported  in part by the National Natural Science Foundation of China under Grant  61722308. ({\em Corresponding author: Keyou You})}
	 \thanks{The authors are with the Department of Automation, and Beijing National Research Center for Info. Sci. \& Tech. (BNRist), Tsinghua University, Beijing 100084, China. E-mail: dongf17@mails.tsinghua.edu.cn, youky@tsinghua.edu.cn.}%
}

\begin{document}

\maketitle
\thispagestyle{empty}
\pagestyle{empty}
\begin{abstract}
The isoline tracking of this work is concerned with the control design for a sensing robot to track a given isoline of an unknown 2-D scalar filed. To this end, we propose a coordinate-free controller with a simple PI-like form using only the concentration feedback for a Dubins robot, which is particularly useful in GPS-denied environments. The key idea lies in the novel design of a sliding surface based error term in the standard PI controller. Interestingly, we also prove that the tracking error can be reduced by increasing the proportion gain, and is eliminated for circular fields with a non-zero integral gain. The effectiveness  of our controller is validated via simulations by using a fixed-wing UAV on the real dataset of the concentration distribution of PM 2.5  in Handan, China.
\end{abstract}


\begin{figure*}[!t] 
	\begin{minipage}[t]{0.34\linewidth}
		\centering{\includegraphics[width=1.0\linewidth]{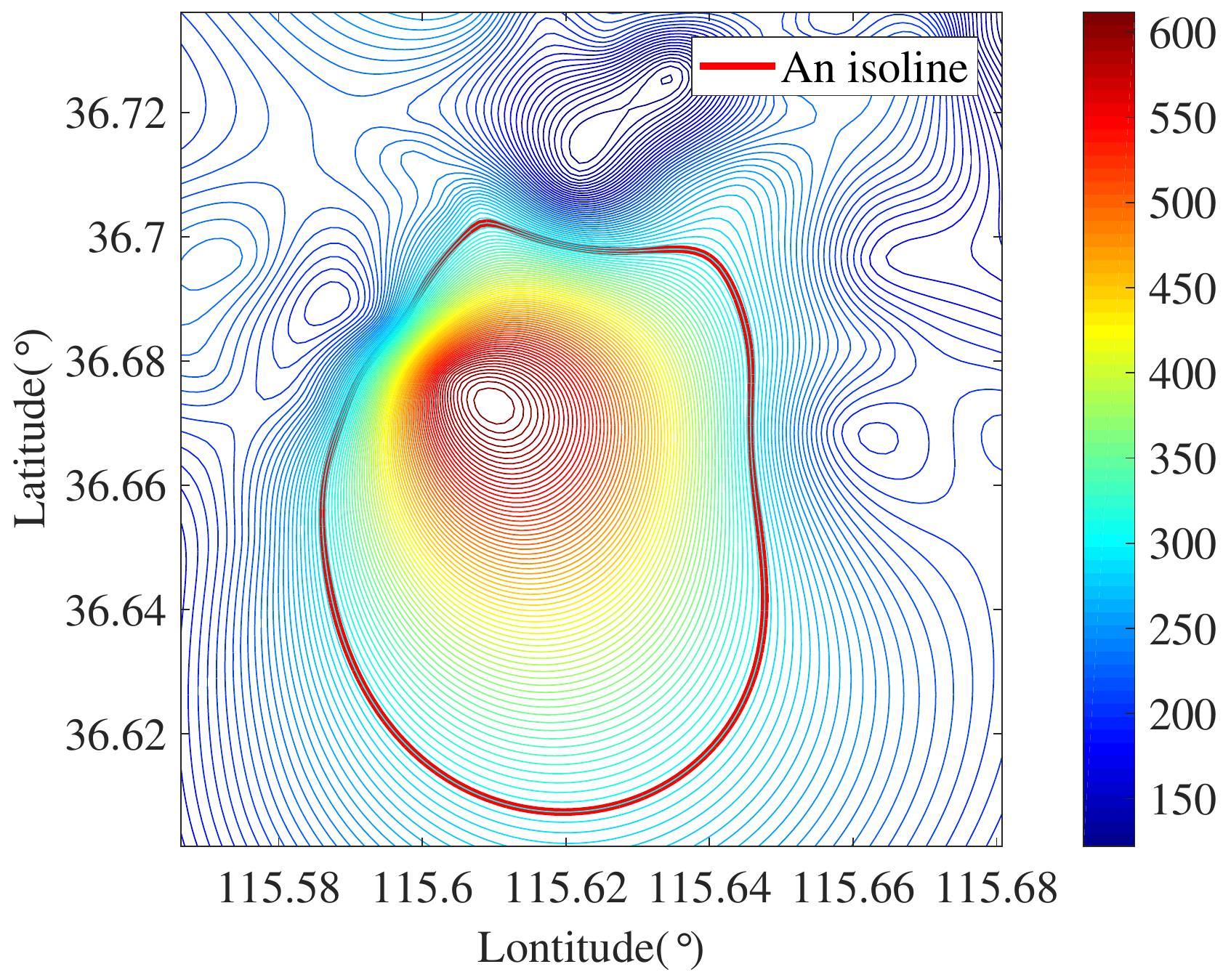}}
		\centerline{(a)}
	\end{minipage}%
	\begin{minipage}[t]{0.33\linewidth}
		\centering{\includegraphics[width=1.0\linewidth]{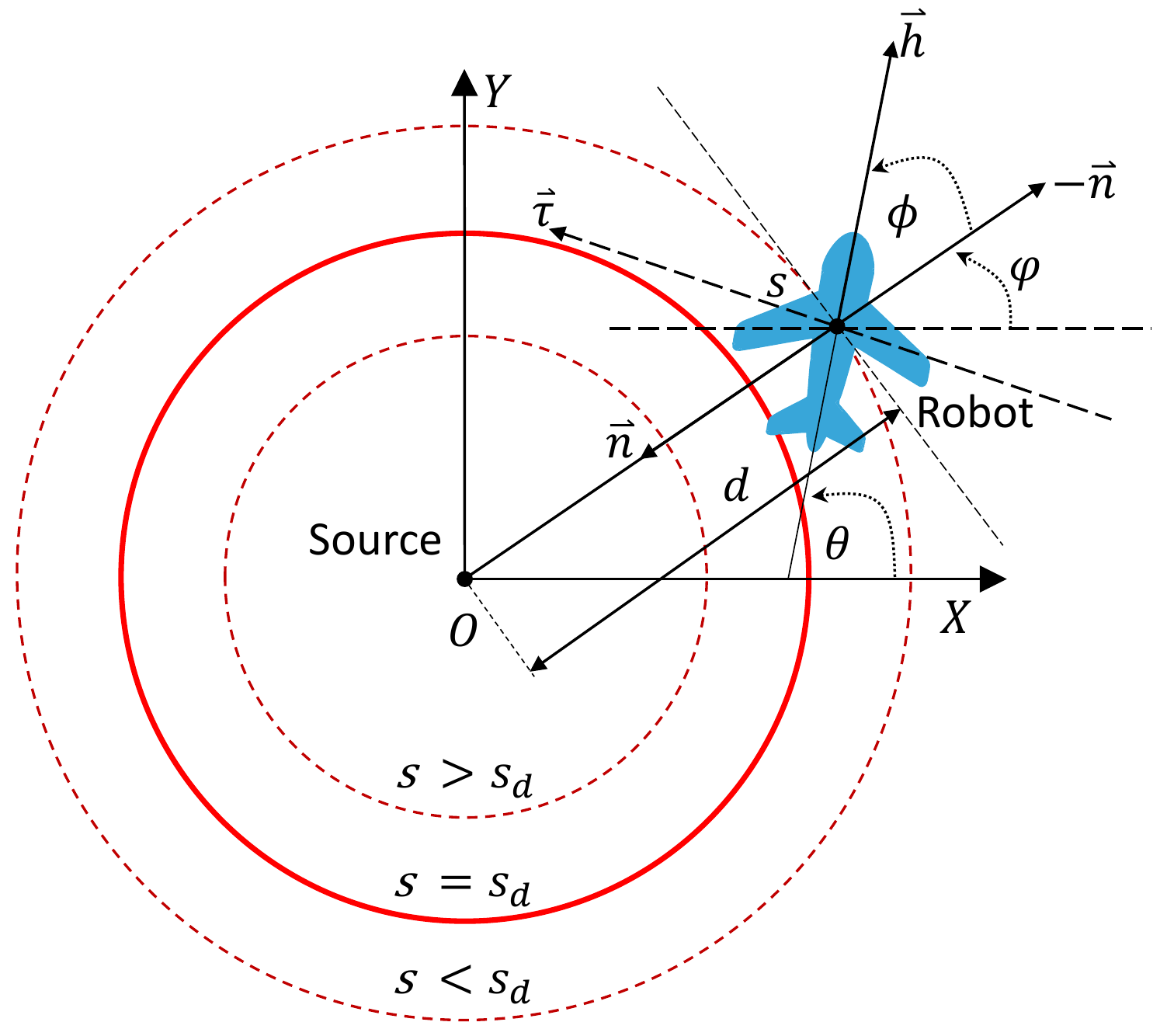}}
		\centerline{(b)}
	\end{minipage}%
	\begin{minipage}[t]{0.33\linewidth}
		\centering{\includegraphics[width=1.0\linewidth]{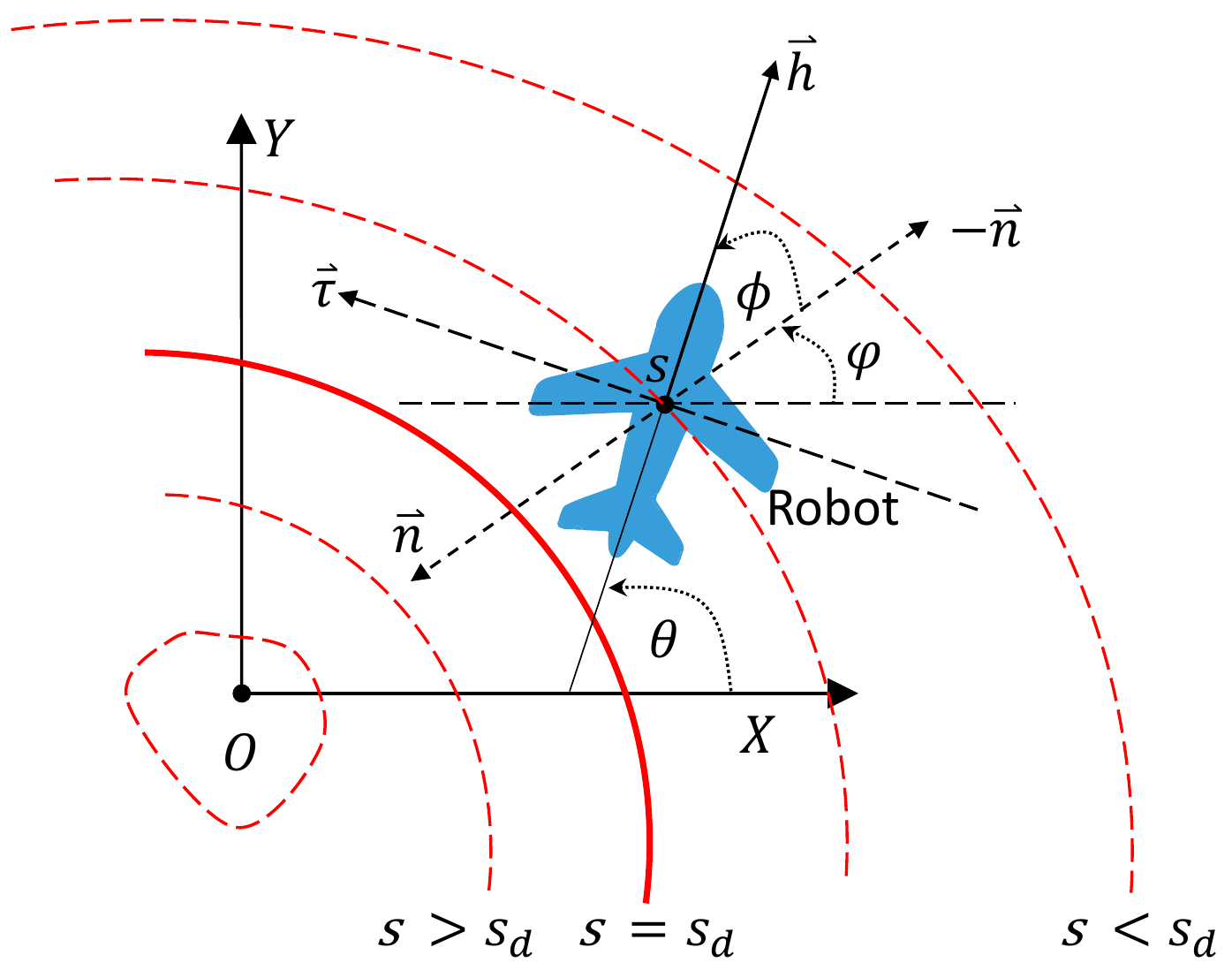}}
		\centerline{(c)}
	\end{minipage}%
		\vspace*{-1.0pt}
	\caption{(a) PM2.5 concentration observed in Handan, a city in North China, at 2018/11/25 22:00:00. (b) Coordinates of the Dubins robot in scalar fields. (c) Coordinates of the Dubins robot in circular fields.}
	\label{fig_illustration}
	\vspace{-0.1cm}	
\end{figure*}

\section{Introduction}
The isoline tracking refers to the tactic that a mobile robot reaches and then tracks a predefined contour in a scalar field, which is widely applied in the areas of detection, exploration, monitoring, and etc. In the literature, it is also named as curve tracking \cite{malisoff2017adaptive}, boundary tracking \cite{matveev2017tight,mellucci2019environmental}, level set tracking \cite{Matveev2012Method}.  In fact, it covers the celebrated target circumnavigation as a special case \cite{deghat2012target,swartling2014collective,dong2020Circumnavigating}.

Compared with the static sensor networks, it is more flexible and economical to utilize mobile sensors to collect data or track target. The methods for isoline tracking by robots have been applied to many practical problems, e.g., exploring environmental feature of bathymetric depth \cite{mellucci2019environmental}, tracking boundary of volcanic ash \cite{kim2017disturbance}, tracking curve of sea temperature \cite{zhang2010cooperative}, and monitoring algal bloom\cite{fonseca2019cooperative}.

Roughly speaking, we can categorize the methods for isoline tracking depending on whether the gradient of the scalar field can be used or not. The gradient-based method is extensively used to the extreme seeking problem, which steers a robot to track the direction of gradient descending (ascending) to reach the minimizer (maximizer) of a scalar field \cite{zhang2010cooperative,wu2012robust}. 

If the explicit gradient is not available, many works focus on the problem of gradient estimation, which mainly include two main strategies: (i) a single robot changes its position over time to collect the signal propagation at different locations; and (ii) multiple robots collaborate to obtain measurements at different locations at the same time.  For the case (i), Ai et al. \cite{ai2016source} show a sequential least-squares field estimation algorithm for a REMUS AUV to seek the source of a hydrothermal plume. Moreover, the stochastic method for extreme seeking is also gradient-based, the idea behind which is to approximate the gradient of the signal strength and to use this information to drive the robot towards the source by adding an excitatory input to the robot steering control \cite{cochran20093,lin2017stochastic}. For the case (ii), a circular formation of robots is adopted in \cite{brinon2015distributed,brinon2019multirobot} to estimate the gradient of fields.  Moreover, a provably convergent cooperative Kalman filter and a cooperative $H_{\infty}$ filter are devised to estimate the gradient in  \cite{zhang2010cooperative} and  \cite{wu2012robust}, respectively.  

In many scenarios, robots cannot obtain its position and can only measure the signal strength at the current location of the sensor, i.e., the measurement in a point-wise fashion \cite{Matveev2012Method}. Thus, it is impossible to estimate the field gradient, and researchers turn to exploiting gradient-free methods.   A sliding mode approach is proposed for target circumnavigation by \cite{Matveev2011Range} and then is adopted to similar problems, e.g., level sets tracking \cite{Matveev2012Method}, boundary tracking \cite{matveev2015robot}, etc. Without a rigorous justification, they address the ``chattering" phenomenon by modeling dynamics of the actuator as the simplest first order linear differential equation in implementation. A PD controller is devised in \cite{baronov2007reactive} for a double-integrator robot to track isolines in a harmonic potential field. Besides, a PID controller with adaptive crossing angle correction is shown in \cite{newaz2018online}. Furthermore, there are some heuristic methods for isoline tracking, e.g., sub-optimal sliding mode algorithm of \cite{mellucci2017experimental}.

In this paper, we propose a coordinate-free controller in a  PI-like from for a Dubins robot to track a desired isoline by using only the concentration feedback. That is, we do not use any field gradient or the position of the robot, which renders our controller particularly useful in the GPS-denied environment. Our key idea lies in the novel design of a  sliding surface based error term in the standard PI controller. 
Similar to the standard PI controller, we show that the final tracking error can be reduced by increasing the proportion gain, and is eliminated for circular fields with a non-zero integral gain.  For the case of smoothing scalar fields, we explicitly show the upper bound of the steady-state tracking error, which can be reduced by increasing the proportional gain. To validate the effectiveness of our controller, we adopt a fixed-wing UAV to track the isoline of the concentration distribution of PM 2.5  in Handan, China. 

The rest of this paper is organized as follows. In Section \ref{sec2}, the problem under consideration is formulated in details. Particularly, we clearly describe the desired isoline tracking pattern. To achieve the objective, we propose a PI-like controller for a Dubins robot in Section \ref{sec_controller}.  In Section \ref{sec_saclar},  we explicitly show the upper bound of the steady-state error in scalar fields. Moreover, we show that the isoline tracking system is locally exponentially stable in Section \ref{sec_cir}. Simulations are performed in Section \ref{secsim}, and some concluding remarks are drawn in Section \ref{sec6}.

\section{Problem Formulation} \label{sec2}
%

 In  Fig.~\ref{fig_illustration}(a), we provide a 2-D example of the concentration distribution of PM 2.5  in Handan, China on November 25, 2018. 
 In the environmental  monitoring, it is fundamentally important to investigate the concentration distribution of air pollutants.  To achieve it, we design a sensing robot to track an isoline of its distribution function.  Mathematically, the concentration of a 2-D scalar field can be described by
\begin{align} \label{eqsca}
F(\bm p): \mathbb{R}^2 \rightarrow \mathbb{R},
\end{align}
where $\bm p \in \mathbb{R}^2$ is the position. Given a concentration level $s_d$, an isoline $\cL(s_d)$ is defined as 
\begin{align} \label{eqset}
\cL(s_d)=\{\bm p | F(\bm p) = s_d   \}.
\end{align}

The {\em isoline tracking} problem is on the design of a controller for a sensing robot to reach a given isoline and maintain on the isoline with a constant speed. That is,   the objective is to asymptotically steer a sensing robot  such that
 \begin{align} \label{eqobj}
 \lim_{t\rightarrow \infty} |s(t)-s_d | \rightarrow 0~ \& ~\|\dot{\bm p}(t)\| = v,
 \end{align}
 where $s(t)= F(\bm p(t))$ is the concentration measurement of the scalar field  at the GPS position $\bm p(t)$ of the robot and $v$ is its constant linear speed.
For a circular field, e.g., acoustic field, then 
\bee F(\bm p)= I_0 \exp(-\varsigma \Vert \bm p - \bm p_o\Vert _2)\label{circufield},\ene 
where $\bm p_o$ is  the source position of the field and $I_0$, $\varsigma$ are unknown parameters. The isoline tracking in \eqref{eqobj} is exactly reduced to the celebrated circumnavigation problem \cite{deghat2012target,swartling2014collective,dong2020Circumnavigating}. 

 In this work, we are interested in the scenario that both the concentration distribution $F(\bm p)$ and the GPS position of the sensing robot are unknown. Moreover, we cannot measure a continuum of the scalar field, which implies that the gradient-based methods \cite{zhang2010cooperative,malisoff2017adaptive,brinon2019multirobot} cannot be applied here. 


 \section{Controller Design} \label{sec_controller}
 In this section, we design a coordinate-free controller in a PI (proportional integral)-like form for a Dubins robot to complete the isoline tracking problem. The key idea lies in the novel design of a sliding surface based error term in the standard PI controller. 

 \subsection{The PI-like controller for a Dubins Robot}
 Consider a Dubins robot on a 2-D plane
  \begin{equation} \label{eq1}
  		\dot{\bm p}(t)  = v \begin{bmatrix}\cos \theta(t) \\ \sin\theta(t)\end{bmatrix},~~\dot \theta(t)= \omega(t) ,
  \end{equation}
 where $\bm p(t) =[x(t),y(t)]'$, $\theta(t)$, $\omega(t)$ and $v$ are the position, heading course, the tunable angular speed and constant linear speed, respectively.
 
To achieve the objective in \eqref{eqobj} by the Dubins robot \eqref{eq1}, we propose a novel PI-like controller 
  \begin{align} \label{eq2}
  \omega(t)  = k_p e(t) + k_i\sigma(t),
  \end{align}
 where $\dot \sigma(t) =e(t)$, $k_p>0$ and $k_i\ge 0$ are the control parameters to be designed. 
 
 Let the tracking error be $\varepsilon(t)=s(t)-s_d$. The major difference of \eqref{eq2} from the standard PI controller lies in the novel design of the following error term 
  \begin{align} \label{eqerror}
e(t) = \dot \varepsilon(t) + c_1 \tanh \left( {\varepsilon(t)}/{c_2}\right),
 \end{align}
where $c_{1,2}>0$ are constant parameters, and $\tanh(\cdot)$ is the standard hyperbolic tangent function to ensure that the selection of the control parameters is independent of the maximum range of the operating space of the controller.  
 In fact, the error term $e(t)$ in \eqref{eqerror} can also be regarded as a sliding surface. For example, once reaching the surface, i.e., $e(t)=0$,  it follows that
 \begin{align*}
 \dot \varepsilon(t) =- c_1 \tanh \left(\varepsilon(t)/c_2 \right),
 \end{align*}
which further implies that $\varepsilon(t)$ will tend to zero with an exponential convergence speed, i.e., the robot will eventually reach the isoline $\cL(s_d)$. 


 Intuitively, the PI-like controller \eqref{eq2} consists of two terms: (i) the proportional term for global stability, and (ii) the integral term to eliminate the steady-state error. Similar to the standard PI controller, the integral coefficient $k_i$ is generally much smaller than the proportional coefficient $k_p$. It is worth mentioning that $c_1$ affects the convergence speed and $c_2$ affects the sensitivity to the tracking error $\varepsilon(t)$.
 
 Clearly, the PI-like controller \eqref{eq2} of this work only uses the concentration measurement $s(t)$ of the scalar field, and is particularly useful in GPS-denied environments.



\subsection{Comparison with the existing methods}
Some related methods to our proposed control laws are (i) the sliding mode controller in \cite{Matveev2012Method}, (ii) the PD controller in \cite{baronov2007reactive}, and (iii) the sliding mode controller with two-sliding motions in \cite{mellucci2019environmental}. The sliding mode approach in \cite{Matveev2012Method} is originally designed for the problem of target circumnavigation \cite{Matveev2011Range} with range-based measurements, and then is adopted to isoline tracking in \cite{Matveev2012Method}. Besides the existence of the chattering phenomenon, their method cannot achieve zero steady-state error even for the task of circumnavigation. In contrast, our PI-like controller \eqref{eq2} is continuous and particularly useful to isoline tracking in circular fields, since the integral part can exactly eliminate the steady-state error. Moreover, the PD feedback controller in \cite{baronov2007reactive} is devised for a double-integrator robot, and their control parameters depend on maximum range of the controller operating space. We address this issue by introducing a hyperbolic tangent function $\tanh(\cdot)$. Furthermore, the controller in \cite{mellucci2019environmental} needs two-sliding motions. They validate their controller by both simulations in a synthetic data-based environment and sea-trials by a C-Enduro ASV in Ardmucknish Bay off Dunstaffnage in Scotland. However, their method is heuristic and in fact only offers uncompleted justification. 

	\newcounter{mytempeqncnt}
	\begin{figure*}[!t]
		\normalsize
		\setcounter{equation}{16}
		\begin{equation} \label{eqnq}
			\hspace{-0.0cm} Q = \bmatri & k_p \alpha v \mu_1^2 -\displaystyle \frac{k_i c_1 \alpha \mu_1}{c_2} & 0 &0 \\ 
			& 0 & (k_p \alpha v)^3 & \displaystyle k_i (k_p \alpha v)^2 - k_i^2 \alpha v /2 + \frac{k_p k_i c_2 (\alpha v)^2 \mu_1}{c_1 \alpha} \\ 
			&0  &\displaystyle k_i (k_p \alpha v)^2 - k_i^2 \alpha v /2 + \frac{k_p k_i c_2 (\alpha v)^2 \mu_1}{c_1 \alpha} & k_p k_i^2 \alpha v \ematri	
		\end{equation}		
		\setcounter{equation}{7}
		\hrulefill	
		\vspace*{0pt}	
	\end{figure*}
	
\section{Isoline Tracking in Circular Fields} \label{sec_cir}
In this section, we first consider the case of a circular field in \eqref{circufield}.  Taking logarithmic function on both sides of  \eqref{circufield}, there is no loss of generality to write it in the following form 
\begin{align}\label{eq_doubleield}
	F(\bm p) = s_d - \alpha ( d(t) - r_d ),
\end{align} 
where $s_d$ is the desired isoline, $\alpha\ge \underline{\alpha}$ is an unknown positive constant, $d(t) = \Vert \bm p(t) - \bm p_o \Vert _2$ is the distance from the robot to the position $\bm p_o$ of the source, and $r_d$ denotes the {\em unknown} radius when the robot travels on the desired isoline, i.e., $s(t) =s_d$.  

Let $\bm n = \nabla F(\bm p)$ denote the gradient vector of $F(\bm p)$, see Fig.~\ref{fig_illustration}(b), and $\bm h =[\cos \theta, ~\sin \theta]'$ represent the course vector of the Dubins robot and $\bm \tau$ to represent the tangent vector of $\bm h$. By convention, $\bm h$ and $\bm \tau$ form a right-handed coordinate frame with $\bm h \times \bm \tau$ pointing to the reader.

After converting the coordinates of the robot from the Cartesian frame into the polar frame, we use the concentration $s(t)$ and angle $\phi(t)$ to describe the tracking system. 
See Fig.~\ref{fig_illustration}(c) for illustrations, where $\bm n$ exactly points to the source and $\phi(t)$ is formed by the negative gradient vector $-\bm n$ and the heading vector $\bm h$. The counter-clockwise direction is set to be positive.


By definitions of $s(t)$ and $\phi(t)$, we have that 
\begin{equation} \label{eqmodel}
	\begin{split}
		\dot s(t) &= -\alpha \dot d(t)=- \alpha v \cos \phi(t), \\
		\dot \phi(t) &= \omega(t) - \frac{v}{d(t)} \sin \phi(t).
	\end{split}
\end{equation}
If $s(t)$ converges to $s_d$, then $d(t)$ also converges to $r_d$. However, $r_d$ is {\em unknown} to the sensing robot, which is substantially different from the target circumnavigation problem \cite{deghat2012target,swartling2014collective}, and we cannot use the control bias $\omega_c = v /r_d$ to eliminate the tracking error as in \cite{dong2020Circumnavigating}.  To solve it, we design an integral term $k_i\sigma(t)$  in \eqref{eq2}.
\prop \label{lemma_circular}
Consider the tracking system in (\ref{eqmodel}) under the PI-like controller in \eqref{eq2}. Define $\bm x(t) = [s(t), ~\phi(t)]'$ and $\bm x_e = [s_d, ~-\pi/2]'$.  If the control parameters are selected to satisfy that
\begin{align}  \label{eqcon2}
k_p(k_p-2) v\underline{\alpha}> k_i~\text{and} ~v\underline{\alpha}>c_1>0,
\end{align} 
then $\bm x_e$ is a locally exponentially stable equilibrium of the tracking system (\ref{eqmodel}).

\begin{proof}
	By \eqref{eqmodel}, the tracking system under the PI-like controller \eqref{eq2} is written as 
	\begin{equation} \label{eq24}
	\begin{split}
	\dot d(t) =&~  v \cos \phi(t), \\
	\dot \phi(t) =&~  - k_p \left(\alpha \dot d(t) +{c_1}\tanh \left({\alpha/c_2\cdot(d(t)-r_d)}\right) \right) \\
	&~+k_i \sigma(t)-{v} \sin \phi(t)/{d(t)},\\
	\dot \sigma(t)  =& -\alpha \dot d(t) + {c_1}\tanh \left({-\alpha/{c_2} \cdot(d(t)-r_d)}\right).
	\end{split}
	\end{equation}
	 Then, we define an error vector
	\begin{align*}
	\bm z(t) &= [z_1(t),~z_2(t),~z_3(t)]' \\
	&= [d(t)-r_d, ~\phi(t)+\pi/2, ~\sigma(t)+v/k_ir_d]' ,
	\end{align*}
	and linearize \eqref{eq24} around $[r_d, ~-\pi/2, ~-v/k_ir_d]'$ as follows
	\begin{align} \label{eqliner}
	\dot {\bm z} (t) = A \bm z(t) ,
	\end{align}	
	where the Jacobian matrix $A$ is given by
	\begin{align*}
	A= \bmatri 0 &  v & 0 \\  -{k_p c_1\alpha}/{c_2 }  - {v}/{r_d^2}  &- {k_p v \alpha} & k_i \\ -c_1\alpha/c_2 & -v\alpha & 0 \ematri.
	\end{align*}
	
	Consider a Lyapunov function candidate as 
	\begin{align} \label{eqvz}
	V(\bm z) =&~ \mu_2 z_1^2(t) + \mu_3 z_2^2(t) + \mu_4 z_3^3(t) \\
	&+ \frac{1}{2}\left( -\mu_1 z_1^2(t) -c_1 v \alpha z_2(t) + c_2 z_3(t)     \right)^2 , \nonumber
	\end{align}
	where $\mu_1 = {k_p c_1\alpha}/{c_2 }  + {v}/{r_d^2}$, $\mu_2 =k_p \alpha (k_p \alpha v \mu_1 - k_i c_1 \alpha/ (2c_2)  ) $, $\mu_3=\mu_1v/2 + k_i \alpha v/2 $, and $\mu_4 =k_p k_i c_2 v \mu_1/c_1 - k_i^2/2 $. It is clear that the conditions in (\ref{eqcon2}) ensure that $V(\bm z)$ is nonnegative.  
	
	Then, we write (\ref{eqvz}) as the following form
	\begin{align}
	V(\bm z) = \bm z' P \bm z,
	\end{align}
	where 
	\begin{align*}
	P =\frac{1}{2} \bmatri 2\mu_2 + \mu_1^2 & k_p \alpha v \mu_1& -k_i \mu_1 \\  k_p \alpha v \mu_1 & 2 \mu_3 +(k_p \alpha v)^2 & -k_p k_i \alpha v \\  -k_i \mu_1 & -k_p k_i \alpha v & 2 \mu_4+ k_i^2   \ematri,
	\end{align*} 	
	which leads to that 
	\begin{align} \label{eqvp}
	\lambda_{\min}(P) \Vert \bm z \Vert_2^2 \le V(\bm z) \le \lambda_{\max}(P) \Vert \bm z \Vert_2^2,
	\end{align}	
	where $\lambda_{\min}(P)$ and $\lambda_{\max}(P)$ denote the minimum and maximum eigenvalues of $P$.
	
	Taking the derivative of $V(\bm z)$ along with \eqref{eqliner} leads to that 
	\begin{align}\label{eqq}
	\dot V(\bm z) = -\bm z' Q \bm z ,
	\end{align}
	where $Q$ is shown in (\ref{eqnq}) and is positive definite by the conditions in (\ref{eqcon2}).

	Then, it follows from  \eqref{eqvp} and \eqref{eqq} that
	\setcounter{equation}{17}
	\begin{equation}
	\dot V(\bm z) \le- \lambda_{\min}(Q) \Vert \bm z \Vert_2^2 \le-  \frac{ \lambda_{\min}(Q) }{\lambda_{\max}(P)} V(\bm z).
	\end{equation}
	By the comparison principle \cite{Khalil2002Nonlinear}, the tracking system \eqref{eqmodel} is locally exponentially stable under the PI-like controller \eqref{eq2}. 
	
\end{proof}

\begin{figure*}[!t]
	\begin{minipage}[t]{0.31\linewidth}
		\centering{\includegraphics[width=1.0\linewidth]{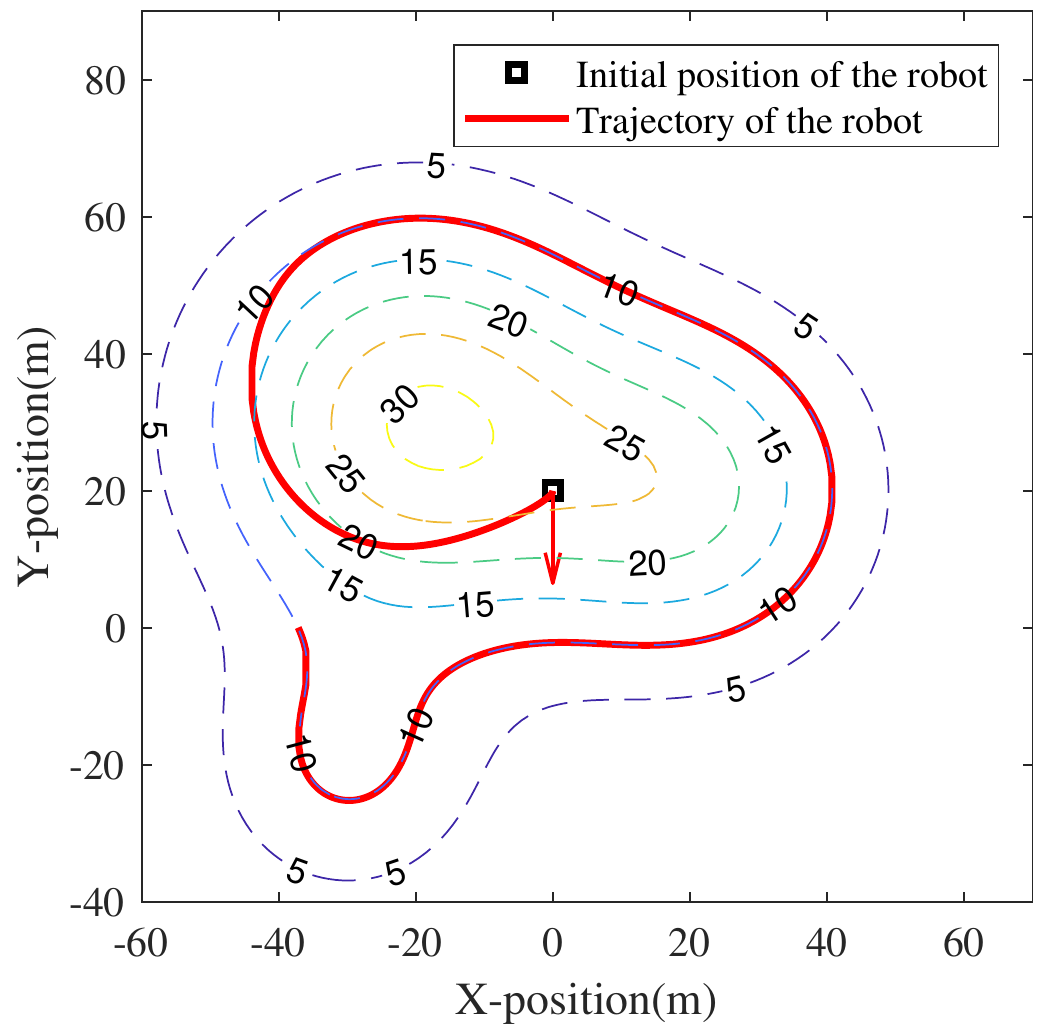}}
		\centerline{(a)}
	\end{minipage}%
	\begin{minipage}[t]{0.31\linewidth}
		\centering{\includegraphics[width=1.0\linewidth]{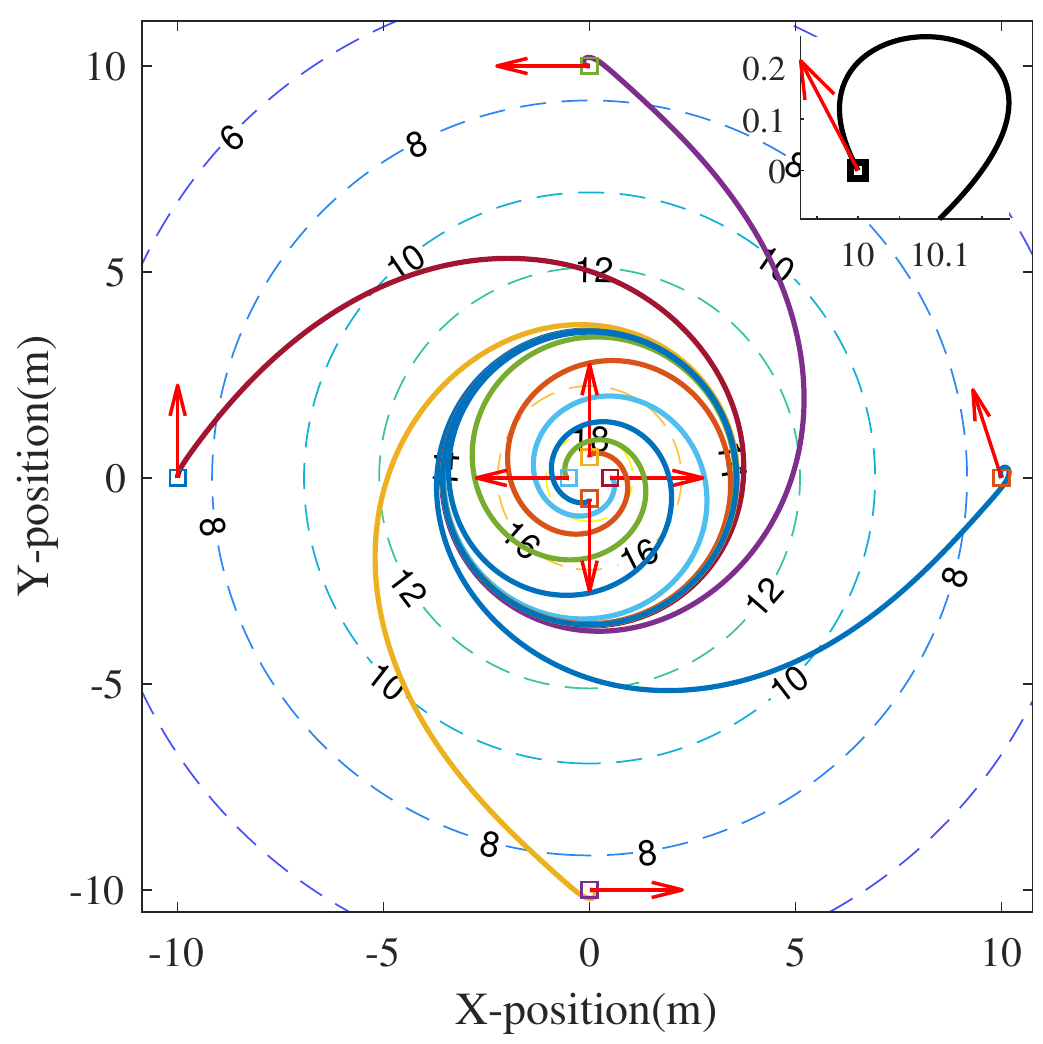}}
		\centerline{(b)}
	\end{minipage}%
	\begin{minipage}[t]{0.38\linewidth}
		\centering{\includegraphics[width=1.0\linewidth]{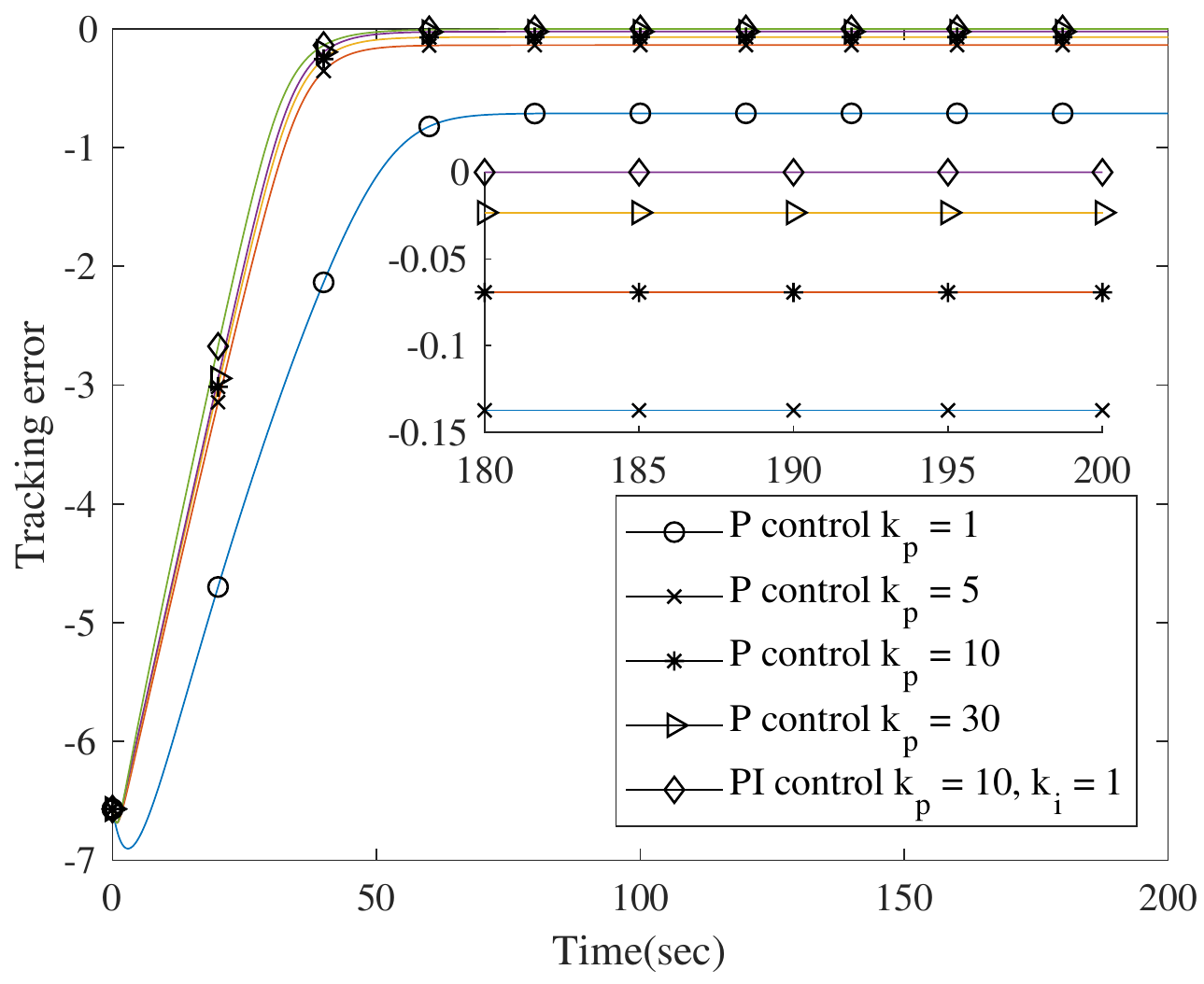}}
		\centerline{(c)}
	\end{minipage}%
	\caption{(a) Fields distribution and trajectory of the Dubins robot. (b) Trajectories of the Dubins robot with different initial states. (c) Tracking errors of the Dubins robot with different control parameters.}
	\label{fig_simulation}
	\vspace{-0.1cm}	
\end{figure*}

\begin{figure*}[!t]
	\begin{minipage}[t]{0.33\linewidth}	
		\centering{\includegraphics[width=1.0\linewidth]{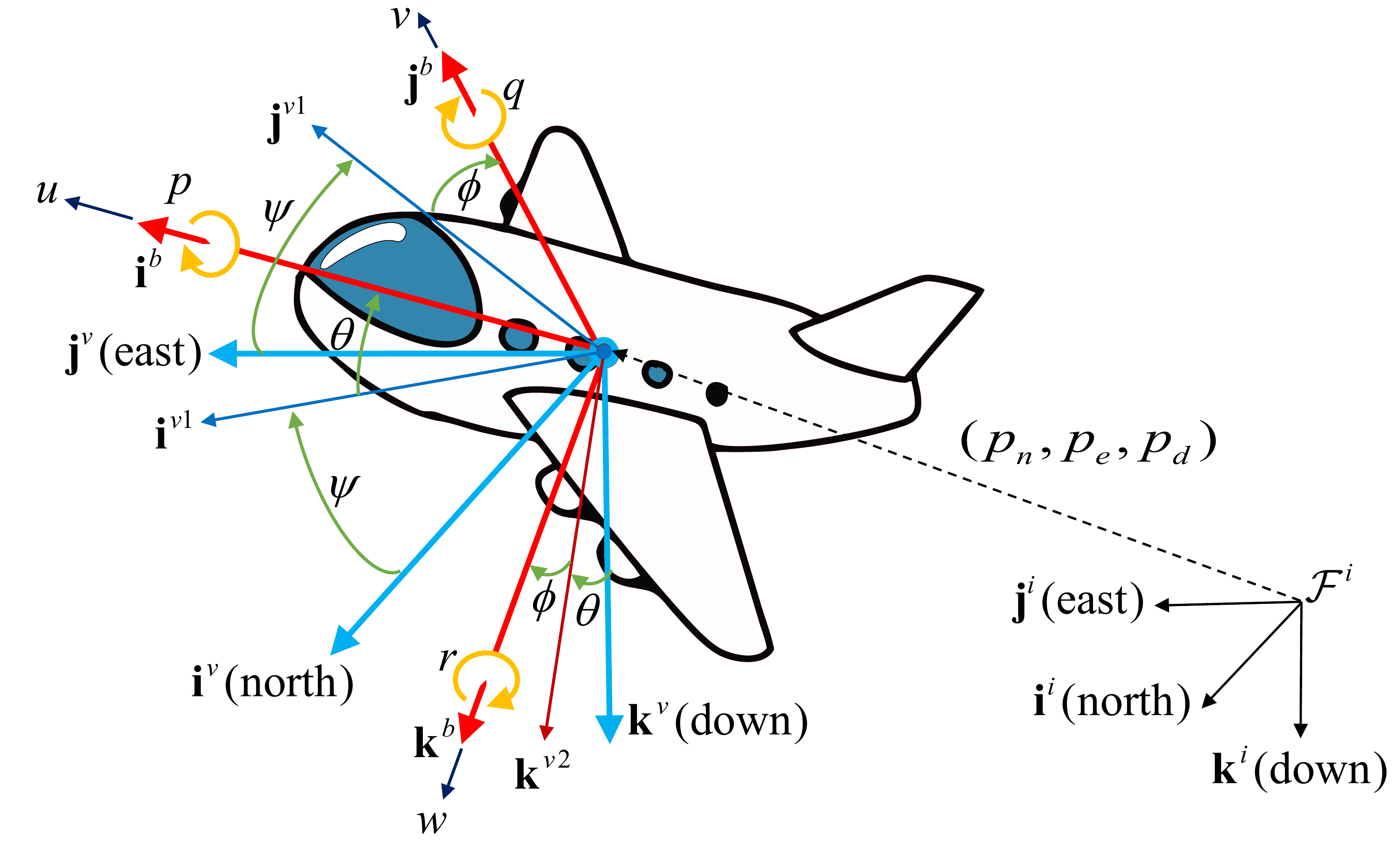}}	
		\centerline{(a)}
	\end{minipage}%
	\begin{minipage}[t]{0.34\linewidth}
	\centering{\includegraphics[width=1.0\linewidth]{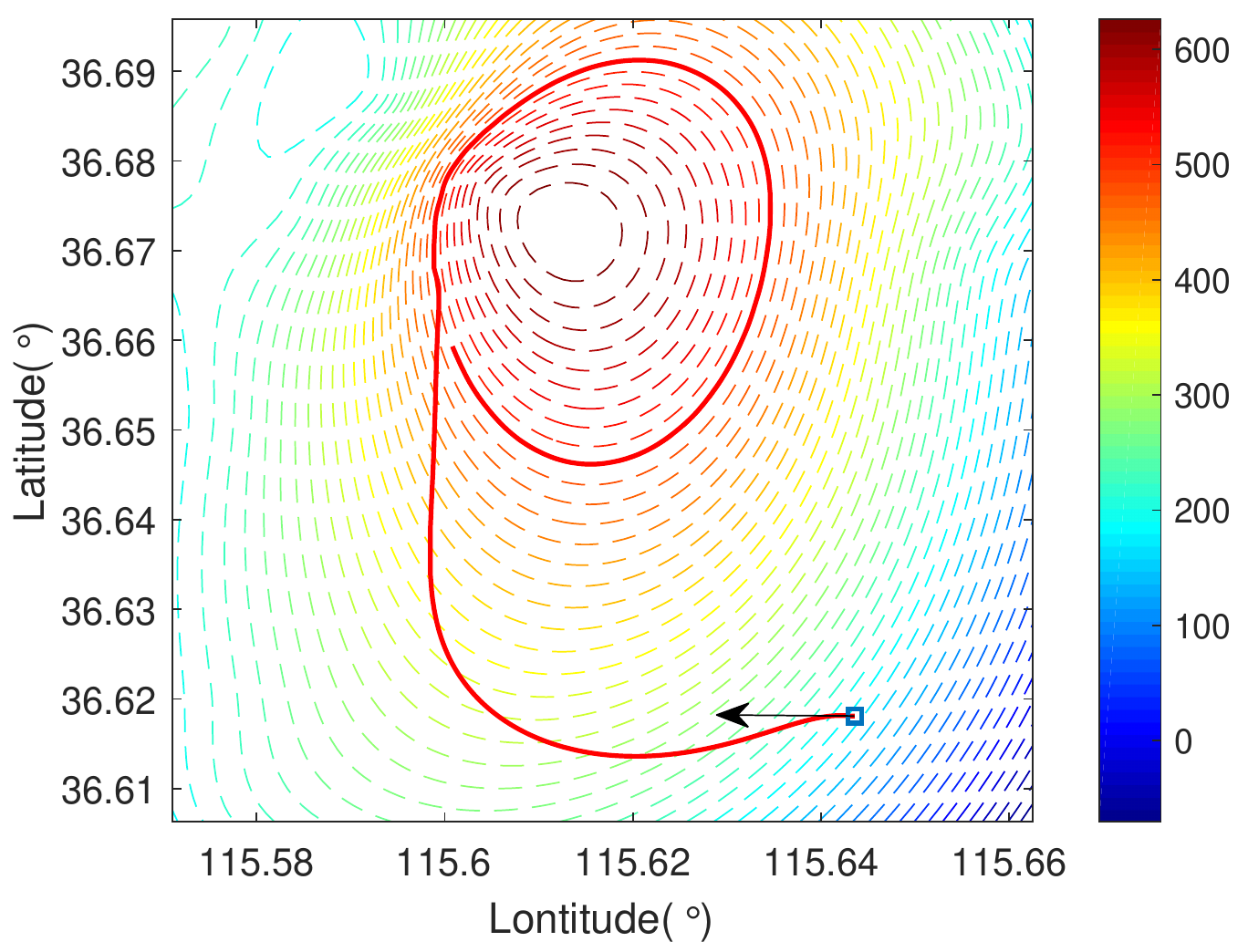}}
		\centerline{(b)}
	\end{minipage}%
	\begin{minipage}[t]{0.33\linewidth}
			\centering{\includegraphics[width=1.0\linewidth]{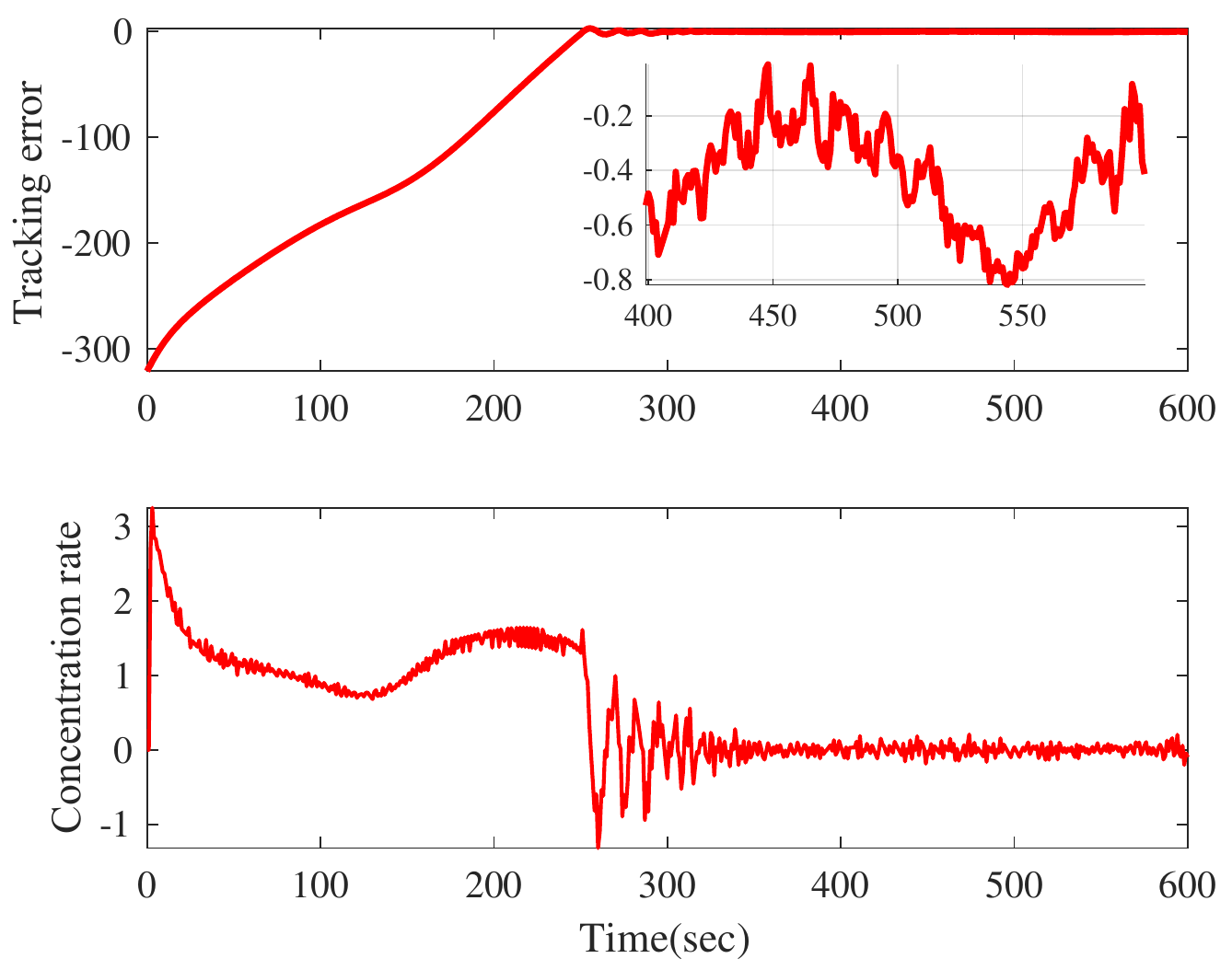}}
		\centerline{(c)}
	\end{minipage}%
	\caption{(a) Coordinates of a fixed-wing UAV. (b) Trajectory of the fixed-wing UAV in the field of PM2.5. (c) Tracking errors and concentration rate of the fixed-wing UAV.}
	\label{fig_fixedwing}
	\vspace{-0.1cm}	
\end{figure*}

\section{Isoline Tracking in Scalar Fields}  \label{sec_saclar}
In this section, we  consider a  scalar field in \eqref{eqsca} under the  assumption  that $F(\bm p)$ is twice differentiable and satisfies 
\begin{align} \label{eqassum}
\gamma_1\le \Vert \nabla F(\bm p) \Vert \le \gamma_2, ~\Vert \nabla^2 F(\bm p) \Vert \le \gamma_3,~ \forall \bm p\in\bR^2
\end{align}
where $\gamma_i$ is a positive constant.  Note from \eqref{eqassum} that $ | \bm h '\nabla^2 F(\bm p) \bm h| \le \gamma_3$ for any $\bm h =[\cos \theta, ~\sin \theta]'$.

To this end, we follow from Fig.~\ref{fig_illustration}(c) that 
\begin{align} \label{eq_dots}
\dot s(t)  =   v \bm n' \bm h = -v \Vert \nabla F(\bm p) \Vert \cos \phi(t).
\end{align}
Then, taking the derivative of $\dot s(t)$ leads to that
\begin{align} \label{eq_ddots}
\ddot s(t) 
&= \omega(t) v \bm n' \bm \tau  +  v^2 \bm h '\nabla^2 F(\bm p) \bm h\\
&= \omega(t) v \Vert \nabla F(p) \Vert \sin \phi(t) +  v^2 \bm h '\nabla^2 F(\bm p) \bm h\nonumber.
\end{align}

\prop \label{prop_quad} Consider the isoline tracking system in \eqref{eq_dots} and \eqref{eq_ddots} under the PI-like controller in \eqref{eq2} and \eqref{eqassum}. If  $\phi(t_0) \in [-\epsilon, -\pi +\epsilon]$ where $\epsilon \in (0,\pi/2)$ and the control parameters are selected to satisfy that
\begin{align*}
k_p > \max\left( \frac{\gamma_3 v}{\gamma_1\sin \epsilon \left ( v \gamma_1 \cos \epsilon - c_1  \right)}, ~ \frac{c_2\gamma_3 v +{c_1\gamma_2} }{  c_1\gamma_1  \sin \epsilon}\right), 
\end{align*}
and $k_i=0$, then 
\begin{align*}
\limsup _{t\rightarrow \infty} |s(t) -s_d |\le \tanh^{-1} \left( \frac{c_2\gamma_3 v +{c_1\gamma_2} }{k_p  c_1\gamma_1  \sin \epsilon}  \right).
\end{align*}
The proof depends on the following technical result. 
 \lemma \label{lemma_bound}
Consider the following system
\begin{align} \label{eq_bound}
\dot z(t) = -k\tanh(z(t)) + b.
\end{align}
If $k>b>0$, then $\limsup _{t \rightarrow \infty}|z(t) | \le \tanh^{-1}\left( {b}/{k}  \right).$
\begin{proof}
	Consider a Lyapunov function candidate as 
	\begin{align*}
	V_z(z) = {1}/{2}\cdot z^2(t).
	\end{align*}
	Taking the derivative of $V_z(z)$ along with \eqref{eq_bound} leads to that
	\begin{align*}
	\dot V_z(z) 
	& = z(t) \left (  -k\tanh(z(t)) + b \right)  \\
	&\le -k z(t) \tanh(z(t)) + b |z(t)|.
	\end{align*}
	By $k>b>0$, it holds that $\dot V_z(z)\le 0$ for all $|z(t)| \ge \tanh^{-1} \left({b}/{k} \right)$. Furthermore,
	it follows that $	\limsup_{t\rightarrow\infty} |z(t)| \le \tanh^{-1} \left({b}/{k} \right).$
\end{proof}

\remark Given a specific $b$ in \eqref{eq_bound}, we can reduce the upper bound by increasing the gain $k$. Similarly,  Proposition \ref{prop_quad} implies that increasing $k_p$ can reduce the upper bound of the steady-state tracking error. 


\begin{proof}[of Proposition \ref{prop_quad}]
	Firstly, we show that $\phi(t)$ can not escape from the region $[-\epsilon, -\pi +\epsilon]$. Substituting the PI-like controller \eqref{eq2} into \eqref{eq_ddots} yields that
	\begin{align} \label{eq_ddots1}
	\ddot s(t) =&~  k_p v \bm n' \bm \tau \left ( \dot \varepsilon(t) + c_1 \tanh \left(  \varepsilon(t)/c_2\right)  \right)  +  v^2 \bm h '\nabla^2 F(\bm p) \bm h .
	\end{align}
	Since $\dot s(t)$ and $\phi(t)$ are continuous with respect to time $t$ by \eqref{eq_dots} and \eqref{eq_ddots1}, we only need to verify the sign of $\ddot s(t)$ when $\phi(t) = -\epsilon$ and $-\pi +\epsilon$. 
	When $\phi(t) = -\epsilon $, it follows from \eqref{eq_ddots1} that
	\begin{align}\label{eqlower} 
	\ddot s(t) = &~ v^2 \bm h '\nabla^2 F(\bm p) \bm h  - k_p v \Vert \nabla F(p) \Vert \sin \epsilon \times\nonumber \\
	&\left ( v \Vert \nabla F(p) \Vert \cos \epsilon + c_1 \tanh \left(  \varepsilon(t)/c_2\right)  \right) \nonumber  \\
	\le&- k_p v \gamma_1\sin \epsilon \left ( v \gamma_1 \cos \epsilon - c_1  \right)  + \gamma_3 v^2  <  0.
	\end{align}
	Similarly, $\phi(t) = -\pi +\epsilon$ yields that 
	\begin{align} \label{equpper}
	\ddot s(t) 
	\ge &- k_p v \gamma_1\sin \epsilon \left ( -v \gamma_1 \cos \epsilon + c_1  \right)  - \gamma_3 v^2  >0.
	\end{align}
	Thus, $\phi(t)$ stays in the region $[-\epsilon,-\pi+\epsilon]$ for all $t\ge t_0$ if $\phi(t_0)\in[-\epsilon,-\pi+\epsilon]$.
	
	Consider a Lyapunov function candidate as 
	\begin{align*}
	V_e(e) = {1}/{2}\cdot e^2(t).
	\end{align*}
	Its derivative along with \eqref{eq_dots} and \eqref{eq_ddots1} is obtained as 
	\begin{align*}
	\dot V_e(e) 
	&= e(t) \left(\ddot s(t)  + {c_1}/{c_2}\cdot \left( 1- \tanh^2\left( \varepsilon(t)/c_2 \right) \right) \dot s(t)   \right)\\
	&=    k_p v \bm n' \bm \tau   e^2(t)  +e(t) \times\\
	&~~~~ \left(  v^2 \bm h '\nabla^2 F(\bm p) \bm h  + {c_1}/{c_2}\cdot \left( 1- \tanh^2\left( \varepsilon(t)/c_2 \right) \right) \dot s(t)   \right) \\
	&\le     k_p v \bm n' \bm \tau   e^2(t)  + \left(\gamma_3 v^2 +{c_1}/{c_2}\cdot \gamma_2 v \right) |e(t)|        \\
	&\le    - \left(k_p v \gamma_1  \sin \epsilon \right)  e^2(t)  + \left(\gamma_3 v^2 +{c_1}/{c_2}\cdot \gamma_2 v \right) |e(t)|.
	\end{align*}
	Thus, $\dot V_e(e)\le 0$ holds for all 
	\begin{align*}
	| e(t)| \ge \rho = \frac{\gamma_3 v +{c_1 \gamma_2}/{c_2}  }{k_p  \gamma_1  \sin \epsilon}.
	\end{align*}
	This implies that $|e(t)|$ will be eventually bounded by $\rho$, i.e., 
	\begin{align*}
	\limsup _{t\rightarrow \infty}\left| \dot \varepsilon(t) + c_1\tanh \left ( {\varepsilon(t)}/{c_2}  \right) \right|  \le \rho .
	\end{align*}
	By Lemma \ref{lemma_bound}, it holds that
	\begin{align*}
	\limsup _{t \rightarrow \infty}|s(t) - s_d| \le \tanh^{-1} \left( \frac{c_2\gamma_3 v +{c_1\gamma_2} }{k_p  c_1\gamma_1  \sin \epsilon}  \right).
	\end{align*}
\end{proof}

\section{Simulations} \label{secsim}
The effectiveness and advantages of the PI-like controller are validated by simulations in this section. Particularly, the PI-like controller \eqref{eq2} is performed on a  realistic simulator of a 6-DOF fixed-wing UAV  \cite{Beard2012Small}.
\subsection{Isoline Tracking in Scalar Fields} \label{sec_sub1}
\begin{table}[!t]
	\caption{Parameters of the controller \eqref{eq2} in Section \ref{sec_sub1}}
	\centering	
	\begin{tabular}{ccccc}	
		\toprule
		{Parameter}   &{$k_p$} &{$k_i$} &{$c_1$} &{$c_2$}\\
		\midrule
		{Value}         & 10          & 0  & 0.1  & 1              \\       
		\bottomrule
	\end{tabular}%
	\label{tab2}%
\end{table}%
Consider a Dubins robot in \eqref{eq1}, and let $\bm q(t) = [\bm p'(t),~\theta(t)]'$ denote its state. The linear speed of the robot is set as $v = 0.5$ \si{m/s}.  Let the Dubins robot travel in a scalar field of Fig.~\ref{fig_illustration}(a),  under the PI-like controller \eqref{eq2} with the parameters shown in Table \ref{tab2}. 
The field distribution and the trajectory of the Dubins robot are given in Fig.~\ref{fig_simulation}(a) with $s_d = 10$ and $\bm q(t_0)=[0,~ 20, ~-\pi/2]$. It is clear that the objective \eqref{eqobj} is eventually achieved.

\subsection{Isoline Tracking in Circular Fields} \label{sec_sub2}
\begin{table}[!t]
	\caption{Parameters of the controller \eqref{eq2} in Section \ref{sec_sub2}}
	\centering	
	\begin{tabular}{ccccc}	
		\toprule
		{Parameter}   &{$k_p$} &{$k_i$} &{$c_1$} &{$c_2$}\\
		\midrule
		{Value}         & 10          & 1  & 0.2  & 1              \\       
		\bottomrule
	\end{tabular}%
	\label{tab3}%
\end{table}%
In this subsection, we validate the performance of the PI-like controller \eqref{eq2} in a circular field 
\begin{align}
F(\bm p) = 20 \exp\left(-0.1 \sqrt{x^2 + y^2}\right)
\end{align}
where the source position is set to origin. The control parameters are selected as Table \ref{tab3}. Fig.~\ref{fig_simulation}(b) illustrates the field distribution and trajectories of the Dubins robot with different initial states.
  Furthermore, Fig.~\ref{fig_simulation}(c) depicts the tracking errors with different control parameters. It can be observed that increasing $k_p$ can exactly enforce the steady-state error to approach zero, however only the controller \eqref{eq2} with $k_i=1$ eventually achieves the objective in \eqref{eqobj} with a zero steady-state error.


\subsection{Isoline Tracking in a field of PM2.5}

In this subsection, a 6-DOF fixed-wing UAV \cite{Beard2012Small} is adopted to test the effectiveness of the PI-like controller \eqref{eq2} in the field of PM2.5,  see  Fig.~\ref{fig_illustration}(a) and Fig.~\ref{fig_fixedwing}(a). To be  consistent with the notions in \cite{Beard2012Small,dong2020Circumnavigating},  we also adopt $[p_n,p_e,p_d]'$ and $[\phi,\theta,\psi]'$ to denote the position and orientation of the UAV in the inertial coordinate frame, respectively. Moreover, we use $[u,v,w]'$ and $[p,q,r]'$ to denote the linear velocities and angular rates in the body frame. Due to page limitation, we omit details of the mathematical model of the UAV, which can be found in \cite{Beard2012Small}, and adopt codes from \cite{small} for the model. Moreover, Fig.~\ref{fig_fixedwing}(b) depicts the distribution of the PM2.5 and the trajectory of the UAV, where the square and arrow denote its initial position and course. Furthermore, the tracking error and the concentration measurement rate of the sensing robot versus time are illustrated in Fig.~\ref{fig_fixedwing}(c). In details, the sampling frequency for the PM2.5 is set as $1$ \si{Hz} and the linear speed of the UAV is maintained as $30$ \si{m/s} by its original controller. 

Overall,  the objective \eqref{eqobj} is eventually achieved by the Dubins robot \eqref{eq1} under the proposed PI-like controllers \eqref{eq2}.

\section{Conclusion} \label{sec6}
To track a desired isoline of a scalar field,  we have designed a coordinate-free controller in a simple PI-like form for a Dubins robot by using concentration-based measurements in this work. A novel idea lies in the design of a sliding surface based error term, which render our PI-like controller different from the standard PI controller. Moreover, the simulation results validated our theoretical finding.

\bibliographystyle{IEEEtran}
\bibliography{bib/mybib}

\begin{thebibliography}{10}
\providecommand{\url}[1]{#1}
\csname url@samestyle\endcsname
\providecommand{\newblock}{\relax}
\providecommand{\bibinfo}[2]{#2}
\providecommand{\BIBentrySTDinterwordspacing}{\spaceskip=0pt\relax}
\providecommand{\BIBentryALTinterwordstretchfactor}{4}
\providecommand{\BIBentryALTinterwordspacing}{\spaceskip=\fontdimen2\font plus
\BIBentryALTinterwordstretchfactor\fontdimen3\font minus
  \fontdimen4\font\relax}
\providecommand{\BIBforeignlanguage}[2]{{%
\expandafter\ifx\csname l@#1\endcsname\relax
\typeout{** WARNING: IEEEtran.bst: No hyphenation pattern has been}%
\typeout{** loaded for the language `#1'. Using the pattern for}%
\typeout{** the default language instead.}%
\else
\language=\csname l@#1\endcsname
\fi
#2}}
\providecommand{\BIBdecl}{\relax}
\BIBdecl

\bibitem{malisoff2017adaptive}
M.~Malisoff, R.~Sizemore, and F.~Zhang, ``Adaptive planar curve tracking
  control and robustness analysis under state constraints and unknown
  curvature,'' \emph{Automatica}, vol.~75, pp. 133--143, 2017.

\bibitem{matveev2017tight}
A.~S. Matveev, A.~A. Semakova, and A.~V. Savkin, ``Tight circumnavigation of
  multiple moving targets based on a new method of tracking environmental
  boundaries,'' \emph{Automatica}, vol.~79, pp. 52--60, 2017.

\bibitem{mellucci2019environmental}
C.~Mellucci, P.~P. Menon, C.~Edwards, and P.~G. Challenor, ``Environmental
  feature exploration with a single autonomous vehicle,'' \emph{IEEE
  Transactions on Control Systems Technology}, 2019.

\bibitem{Matveev2012Method}
A.~S. Matveev, H.~Teimoori, and A.~V. Savkin, ``Method for tracking of
  environmental level sets by a unicycle-like vehicle,'' \emph{Automatica},
  vol.~48, no.~9, pp. 2252---2261, 2012.

\bibitem{deghat2012target}
M.~Deghat, E.~Davis, T.~See, I.~Shames, B.~D. Anderson, and C.~Yu, ``Target
  localization and circumnavigation by a non-holonomic robot,'' in \emph{2012
  IEEE/RSJ International Conference on Intelligent Robots and Systems
  (IROS)}.\hskip 1em plus 0.5em minus 0.4em\relax Vilamoura: IEEE, 2012, pp.
  1227--1232.

\bibitem{swartling2014collective}
J.~O. Swartling, I.~Shames, K.~H. Johansson, and D.~V. Dimarogonas,
  ``Collective circumnavigation,'' \emph{Unmanned Systems}, vol.~2, no.~03, pp.
  219--229, 2014.

\bibitem{dong2020Circumnavigating}
F.~Dong, K.~You, and L.~Xie, ``Circumnavigating a moving target with range-only
  measurements,'' \emph{arXiv:2002.06507}, 2020.

\bibitem{kim2017disturbance}
J.-S. Kim, P.~P. Menon, J.~Back, and H.~Shim, ``Disturbance observer based
  boundary tracking for environment monitoring,'' \emph{Journal of Electrical
  Engineering \& Technology}, vol.~12, no.~3, pp. 1299--1306, 2017.

\bibitem{zhang2010cooperative}
F.~Zhang and N.~E. Leonard, ``Cooperative filters and control for cooperative
  exploration,'' \emph{IEEE Transactions on Automatic Control}, vol.~55, no.~3,
  pp. 650--663, 2010.

\bibitem{fonseca2019cooperative}
J.~Fonseca, J.~Wei, K.~H. Johansson, and T.~A. Johansen, ``Cooperative
  decentralized circumnavigation with application to algal bloom tracking,'' in
  \emph{IEEE/RSJ International Conference on Intelligent Robots and Systems
  (IROS)}.\hskip 1em plus 0.5em minus 0.4em\relax Macau, China: IEEE, 2019, pp.
  3276--3281.

\bibitem{wu2012robust}
W.~Wu and F.~Zhang, ``Robust cooperative exploration with a switching
  strategy,'' \emph{IEEE Transactions on Robotics}, vol.~28, no.~4, pp.
  828--839, 2012.

\bibitem{ai2016source}
X.~Ai, K.~You, and S.~Song, ``A source-seeking strategy for an autonomous
  underwater vehicle via on-line field estimation,'' in \emph{14th
  International Conference on Control, Automation, Robotics and Vision
  (ICARCV)}.\hskip 1em plus 0.5em minus 0.4em\relax IEEE, 2016, pp. 1--6.

\bibitem{cochran20093}
J.~Cochran, A.~Siranosian, N.~Ghods, and M.~Krstic, ``{3-D} source seeking for
  underactuated vehicles without position measurement,'' \emph{IEEE
  Transactions on Robotics}, vol.~25, no.~1, pp. 117--129, 2009.

\bibitem{lin2017stochastic}
J.~Lin, S.~Song, K.~You, and M.~Krstic, ``Stochastic source seeking with
  forward and angular velocity regulation,'' \emph{Automatica}, vol.~83, pp.
  378--386, 2017.

\bibitem{brinon2015distributed}
L.~Bri{\~n}{\'o}n-Arranz, L.~Schenato, and A.~Seuret, ``Distributed source
  seeking via a circular formation of agents under communication constraints,''
  \emph{IEEE Transactions on Control of Network Systems}, vol.~3, no.~2, pp.
  104--115, 2015.

\bibitem{brinon2019multirobot}
L.~Bri{\~n}{\'o}n-Arranz, A.~Renzaglia, and L.~Schenato, ``Multirobot symmetric
  formations for gradient and {Hessian} estimation with application to source
  seeking,'' \emph{IEEE Transactions on Robotics}, vol.~35, no.~3, pp.
  782--789, 2019.

\bibitem{Matveev2011Range}
A.~S. Matveev, H.~Teimoori, and A.~V. Savkin, ``Range-only measurements based
  target following for wheeled mobile robots,'' \emph{Automatica}, vol.~47,
  no.~1, pp. 177--184, 2011.

\bibitem{matveev2015robot}
A.~S. Matveev, M.~C. Hoy, K.~Ovchinnikov, A.~Anisimov, and A.~V. Savkin,
  ``Robot navigation for monitoring unsteady environmental boundaries without
  field gradient estimation,'' \emph{Automatica}, vol.~62, pp. 227--235, 2015.

\bibitem{baronov2007reactive}
D.~Baronov and J.~Baillieul, ``Reactive exploration through following isolines
  in a potential field,'' in \emph{American Control Conference}.\hskip 1em plus
  0.5em minus 0.4em\relax IEEE, 2007, pp. 2141--2146.

\bibitem{newaz2018online}
A.~A.~R. Newaz, S.~Jeong, and N.~Y. Chong, ``Online boundary estimation in
  partially observable environments using a {UAV},'' \emph{Journal of
  Intelligent \& Robotic Systems}, vol.~90, no. 3-4, pp. 505--514, 2018.

\bibitem{mellucci2017experimental}
C.~Mellucci, P.~P. Menon, C.~Edwards, and P.~Challenor, ``Experimental
  validation of boundary tracking using the suboptimal sliding mode
  algorithm,'' in \emph{American Control Conference (ACC)}.\hskip 1em plus
  0.5em minus 0.4em\relax IEEE, 2017, pp. 4878--4883.

\bibitem{Khalil2002Nonlinear}
H.~K. Khalil, \emph{Nonlinear Systems (3rd Ed.)}.\hskip 1em plus 0.5em minus
  0.4em\relax Prentice Hall, 2002.

\bibitem{Beard2012Small}
R.~W. Beard and T.~W. Mclain, \emph{Small Unmanned Aircraft: Theory and
  Practice}.\hskip 1em plus 0.5em minus 0.4em\relax Princeton University Press,
  2012.

\bibitem{small}
\BIBentryALTinterwordspacing
J.~Lee, ``Small fixed wing {UAV} simulator,'' Apr. 2016. [Online]. Available:
  \url{https://github.com/magiccjae/ecen674}
\BIBentrySTDinterwordspacing

\end{thebibliography}

\end{document}